\documentclass{article}
\usepackage{bezier,emlines2,gdgspace}
\advance\textheight by -\headsep\advance\textheight by -\headheight
\title{The Three-Box ``Paradox'' and other Reasons to
Reject the Counterfactual Usage of the ABL Rule}
\author{ R.E. Kastner\thanks{\normalsize rkastner@wam.umd.edu}}
\date{Version of February 8, 1999}
\begin{document}
\maketitle
\begin{center}
{\small \em Department of Philosophy \\
University of Maryland \\
College Park, MD 20742 USA. \\}
\end{center}
\vspace{.1cm}
\begin{abstract}
An apparent paradox proposed by Aharonov and Vaidman in
which a single particle can be found with certainty in
two (or more) boxes is analyzed by way of a simple
thought experiment. It is found that the apparent
paradox arises from an invalid counterfactual usage of the Aharonov-Bergmann-Lebowitz
(ABL) rule, and effectively attributes conflicting properties
not to the same particle but to different particles. 
A connection is made between the present analysis
and the consistent histories formulation of Griffiths. Finally,
a critique is given of some recent counterarguments by Vaidman against the rejection
of the counterfactual usage of the ABL rule.
\end{abstract}

\large \vskip .5cm
{\bf 1. Background.~~ }

\def\psizero{\vert\psi_0\rangle}
\def\psizerop{\vert\psi_{0}'\rangle}
\def\psizeropp{\vert\psi_{0}''\rangle}
\def\psif{\vert\psi_f\rangle}
\def\statea{\vert a\rangle}
\def\stateb{\vert b\rangle}
\def\statec{\vert c\rangle}
\def\stateapr{\vert a'\rangle}
\def\statebpr{\vert b'\rangle}
\def\proja{\vert a\rangle\langle a \vert}
\def\projb{\vert b\rangle\langle b \vert}
\def\projc{\vert c\rangle\langle c \vert}
\large \vskip .5cm
The Aharonov-Bergmann-Lebowitz (ABL) rule is a well-known
formula for calculating the probabilities of the possible outcomes
of observables measured at an intermediate time $t_1$ between
pre- and post-selection measurements at times $t_0$ and $t_f$,
respectively [Aharonov, Bergmann, and Lebowitz, 1964]. If
an intermediate measurement of (possibly degenerate) observable
$C$ with
 eigenvalues $\{c_i\}$ is performed at time $t_1$,
the ABL rule states that the probability of outcome $c_j$ in the 
case of a preselection
for the state $\psizero$ and a post-selection for the state 
$\psif$ is given by:\footnote{\normalsize In this case,
the Hamiltonian $H = 0$; the ABL rule also applies to
the more general case of nonzero Hamiltonian, with
a time-dependence of the pre- and post-selection states
in the usual way.}

	$$ P_{ABL}(c_j|\psi_0,\psi_f) = 
{ |\langle \psi_f|P_{c_j}\psizero|^2 \over
{\sum_i |\langle \psi_f |P_{c_i}
\psizero |^2}}   \eqno(1)$$

\noindent where $P_{c_j}$ is the projection operator
on the eigenspace corresponding to outcome $c_j$. 

In the formulation of quantum theory known 
as ``Time Symmetrized Quantum Theory,'' (cf. 
Aharonov and Vaidman, 1991), a system thus pre- 
and post-selected is labelled by a time-symmetric
``two-state vector'' $\Psi = \langle \psi_f | \psizero$.

It has recently been shown\footnote{\normalsize 
Sharp and Shanks (1993), Cohen (1995), Miller (1996),
Kastner (1998).}
 that the ABL rule cannot be used
in a counterfactual sense: i.e., in general, it cannot be used
to calculate the probabilities of possible outcomes
of observables that have not actually been measured at time $t_1$.
This result has been implicit all along in the consistent
histories formulation of Griffiths (1984, 1996, 1998), which
gives precise conditions under which meaningful, ``classical''
probabilities can be assigned to the outcomes of real or hypothetical
measurements. (Indeed, the ABL rule can be seen as one
simple instance of the consistent histories approach; 
this is shown in the Appendix.)

A quantitative distinction is made in Kastner (1998) between
the valid, non-counterfactual usage and the generally
invalid, counterfactual usage. This involves augmenting
the ABL probability $P_{ABL}$ in (1) with an additional
parameter specifying which observable has {actually
been measured in the selection of any
particular system. Thus if system X has been pre-
and post-selected in the state $\Psi = \langle \psi_f | \psizero$
 via an intervening
measurement of observable $C$ at time $t_1$, the augmented
 ABL probability is written as:

$$ P_{ABL}(x_j|\psi_0,\psi_f;C) = 
{ |\langle \psi_f|P_{x_j}
\psizero|^2 \over{\sum_i |\langle \psi_f |P_{x_i}\psizero |^2}} ,  \eqno(1')$$

\noindent where the outcome whose probability is being calculated
is denoted by the general parameter $x_j$. The correct,
non-counterfactual usage restricts $x_j$ to the set
of eigenvalues $\{c_i\}$ of $C$;
the (generally) incorrect, counterfactual usage consists of allowing 
$x_j$ to vary over values not in the
range of $C$.

{\bf 2. The three-box example.}

	In his `Weak-Measurement Elements of Reality' (1996), Vaidman
 discusses an example of what he terms an ``ideal-measurement element 
of reality of the pre- and post-selected system'' (`ideal'
in the sense of ideal measurements
rather than weak measurements; cf. Vaidman (1996, p. 899)). This consists of 
a single particle and three boxes labeled A, B, and C. The particle 
is pre-selected at time $t_0$ in the state
	
$$\psizero = {1\over \sqrt 3} \biggl(\statea + \stateb + \statec\biggr)
\eqno(2a)$$
 
and post-selected at time $t_f > t_0$ in the state

$$\psif = {1\over \sqrt 3} \biggl(\statea + \stateb - \statec\biggr),
\eqno(2b)$$

\noindent where the states $\statea$,  $\stateb$, and $\statec$ correspond
to the particle being found in box A, B, or C, respectively.
 The pre- and 
post-selected states $\psizero$  and $\psif$ are not orthogonal and can be 
viewed as eigenvectors of two different observables $Q_0$
 and $Q_f$. Let the eigenbasis of observable $Q_0$
 be labelled as $\{\psizero, \psizerop, \psizeropp \}$,
and similarly for $Q_f$.

The two possible intermediate
 measurements at time $t_1$ are of two observables $A$ and $B$, 
corresponding to opening box A or opening box B. They are 
defined as follows:

	$$A= a \proja + a' [\projb + \projc],\eqno(3a)$$  and
	$$B= b \projb + b' [\proja + \projc],\eqno(3b)$$

\noindent where $a'$ and $b'$ are the eigenvalues corresponding to finding 
the particle to be {\it not} in box A or B, respectively. 
Their associated eigenspaces are the planes $bc$ and $ac$
(See Figure 1). We also define the states 
$$\vert a'\rangle = {1\over\sqrt 2}(\stateb + \statec)\eqno(4a)$$
$$\vert b'\rangle = {1\over\sqrt 2}(\statea + \statec)\eqno(4b)$$
\noindent obtained by projecting the initial state $\psizero$
onto the degenerate eigenspaces corresponding to
the planes $bc$ and $ac$, respectively. These states describe
the system after a minimally disturbing measurement of $A$ or $B$
yielding outcome $a'$ or $b'$, i.e., the outcome in which the particle
is found to be {\it not} in the box which was opened. 

\hrule%TexCad Options
%\grade{\on}
%\emlines{\on}
%\beziermacro{\off}
%\reduce{\on}
%\snapping{\off}
%\quality{2.00}
%\graddiff{0.01}
%\snapasp{1}
%\zoom{1.00}
\special{em:linewidth 0.1pt}
\unitlength 1.00mm
\linethickness{0.1pt}
\begin{picture}(106.33,150.67)
\special{em:linewidth 0.1pt}
\unitlength 1.00mm
\linethickness{0.1pt}
\emline{56.33}{103.67}{1}{85.00}{103.67}{2}
\emline{85.00}{103.67}{3}{85.00}{46.67}{4}
\emline{85.00}{46.67}{5}{56.67}{46.67}{6}
\emline{56.67}{46.67}{7}{56.67}{76.33}{8}
\emline{64.00}{96.33}{9}{35.00}{96.33}{10}
\emline{35.00}{96.33}{11}{56.33}{103.33}{12}
\emline{63.33}{96.33}{13}{84.67}{103.00}{14}
\emline{64.00}{38.67}{15}{85.33}{46.67}{16}
\emline{63.67}{96.33}{17}{63.67}{38.67}{18}
\emline{63.67}{38.67}{19}{34.67}{38.67}{20}
\emline{34.67}{38.67}{21}{34.67}{96.67}{22}
\emline{35.00}{38.67}{23}{56.33}{46.67}{24}
\emline{35.00}{69.33}{25}{63.67}{69.33}{26}
\emline{63.67}{69.33}{27}{85.00}{76.33}{28}
\special{em:linewidth 0.4pt}
\unitlength 1.00mm
\linethickness{0.4pt}
%\vector(56.33,76.67)(95.00,112.33)
\put(95.00,112.33){\vector(1,1){0.2}}
\emline{56.33}{76.67}{29}{95.00}{112.33}{30}
%\end
%\vector(56.33,76.33)(25.33,106.00)
\put(25.33,106.00){\vector(-1,1){0.2}}
\emline{56.33}{76.33}{31}{25.33}{106.00}{32}
%\end
%\vector(56.33,76.00)(63.33,96.33)
\put(63.33,96.33){\vector(1,3){0.2}}
\emline{56.33}{76.00}{33}{63.33}{96.33}{34}
%\end
%\vector(56.67,76.67)(63.33,38.67)
\put(63.33,38.67){\vector(1,-4){0.2}}
\emline{56.67}{76.67}{35}{63.33}{38.67}{36}
%\end
%\vector(56.33,76.67)(56.33,127.00)
\put(56.33,127.00){\vector(0,1){0.2}}
\emline{56.33}{76.67}{37}{56.33}{127.00}{38}
%\end
%\vector(56.67,76.67)(102.00,76.67)
\put(102.00,76.67){\vector(1,0){0.2}}
\emline{56.67}{76.67}{39}{102.00}{76.67}{40}
%\end
%\vector(56.33,77.00)(16.33,63.33)
\put(16.33,63.33){\vector(-3,-1){0.2}}
\emline{56.33}{77.00}{41}{16.33}{63.33}{42}
%\end
\put(56.33,140.67){\makebox(0,0)[cc]{\normalsize Figure 1}}
\put(56.00,19.33){\makebox(0,0)[cc]
{\normalsize The pre- and post-selected states $|\psi_0\rangle$ and $|\psi_f\rangle$ in the three-box basis.}}
\put(97.00,116.33){\makebox(0,0)[cc]
{$\vert a'\rangle = {1\over\sqrt 2}(|b\rangle + |c\rangle)$}}
\put(52.67,131.00){\makebox(0,0)[cc]{$|c\rangle$}}
\put(10.67,59.00){\makebox(0,0)[cc]{$|a\rangle$}}
\put(106.33,76.67){\makebox(0,0)[cc]{$|b\rangle$}}
\put(60.67,99.67){\makebox(0,0)[cc]{$|\psi_0\rangle$}}
\put(68.00,35.33){\makebox(0,0)[cc]{$|\psi_f\rangle$}}
\special{em:linewidth 0.1pt}
\unitlength 1.00mm
\linethickness{0.1pt}
\emline{60.00}{79.67}{43}{61.00}{74.33}{44}
\emline{61.00}{74.33}{45}{57.67}{72.67}{46}
\emline{53.00}{79.33}{47}{54.33}{73.00}{48}
\emline{54.33}{73.00}{49}{58.00}{69.67}{50}
\put(23.33,110.33){\makebox(0,0)[cc]
{$|b'\rangle={1\over\sqrt 2}(|a\rangle + |c\rangle )$}}
\end{picture}
\hrule\newpage

With the above pre- and post-selection, the ABL rule (1) gives probability one 
for an outcome of either $a$ or $b$
 at time $t_1$ ($t_0 < t_1 < t_f$) upon measurement of observable
$A$ or $B$ corresponding to opening box A or box B.

{\bf 3. Analysis and resolution.}

Vaidman interprets the above results as indicating that 
there are two `elements of reality' for this system, corresponding to both 
the particle being in box A (if we look for it there) and the particle 
being in box B (if we look for it there). These `elements of reality' are, 
indeed, highly peculiar and counterintuitive. But need 
we really accept them as `elements of reality'? I will argue in the 
negative: these results cannot be interpreted as applying to an 
individual system such as the particle in the above example.

	Consider first an experiment in
 which we start 
with a large number of systems pre-selected in 
state $\psizero$ and we choose to 
open box A at time $t_1$, thus measuring observable $A$ \nolinebreak
\hbox{(see Figure 2(a)).} \newpage
%TexCad Options
%\grade{\on}
%\emlines{\on}
%\beziermacro{\off}
%\reduce{\on}
%\snapping{\off}
%\quality{2.00}
%\graddiff{0.01}
%\snapasp{1}
%\zoom{1.00}
\special{em:linewidth 0.4pt}
\unitlength 1.00mm
\linethickness{0.4pt}
\begin{picture}(117.67,100.33)(0,20)
\emline{27.00}{86.67}{1}{64.00}{99.67}{2}
\emline{28.00}{86.67}{3}{64.33}{76.00}{4}
\emline{82.00}{99.67}{5}{107.33}{110.33}{6}
\emline{82.33}{99.67}{7}{107.33}{99.67}{8}
\emline{82.33}{99.67}{9}{107.33}{90.00}{10}
\emline{82.33}{75.67}{11}{107.67}{75.67}{12}
\emline{82.00}{75.33}{13}{107.67}{65.00}{14}
\put(71.33,127.33){\makebox(0,0)[cc]{Figure 2}}
\put(21.00,86.34){\makebox(0,0)[cc]{$|\psi_0\rangle$}}
\put(69.00,100.00){\makebox(0,0)[cc]{$|a\rangle$}}
\put(69.00,75.33){\makebox(0,0)[cc]{$|a'\rangle$}}
\put(75.33,100.33){\makebox(0,0)[cc]{$\bigl( {1\over 3}\bigr)$}}
\put(75.00,75.34){\makebox(0,0)[cc]{$\bigl( {2\over 3}\bigr)$}}
\put(111.33,111.00){\makebox(0,0)[cc]{$|\psi_f\rangle$}}
\put(111.33,99.67){\makebox(0,0)[cc]{$|\psi_{f}'\rangle$}}
\put(111.33,89.67){\makebox(0,0)[cc]{$|\psi_{f}''\rangle$}}
\put(112.33,75.33){\makebox(0,0)[cc]{$|\psi_{f}'\rangle$}}
\put(111.67,64.00){\makebox(0,0)[cc]{$|\psi_{f}''\rangle$}}
\put(117.67,111.00){\makebox(0,0)[cc]{$\bigl( {1\over 9}\bigr)$}}
%\put(113.00,-10.00){\makebox(0,0)[cc]{$|\psi_{f}''\rangle$}}
\put(2.00,86.33){\makebox(0,0)[cc]{(a)}}
\emline{82.00}{75.33}{15}{108.33}{84.67}{16}
\put(117.67,84.33){\makebox(0,0)[cc]{$(0)$}}
\put(111.33,84.33){\makebox(0,0)[cc]{$|\psi_f\rangle$}}
\end{picture}
%TexCad Options
%\grade{\on}
%\emlines{\on}
%\beziermacro{\off}
%\reduce{\on}
%\snapping{\off}
%\quality{2.00}
%\graddiff{0.01}
%\snapasp{1}
%\zoom{1.00}
\special{em:linewidth 0.4pt}
\unitlength 1.00mm
\linethickness{0.4pt}
\begin{picture}(117.67,100.00)(-5,-20)
\emline{27.00}{86.67}{1}{64.00}{99.67}{2}
\emline{28.00}{86.67}{3}{64.33}{76.00}{4}
\emline{82.00}{99.67}{5}{107.33}{110.33}{6}
\emline{82.33}{99.67}{7}{107.33}{99.67}{8}
\emline{82.33}{99.67}{9}{107.33}{90.00}{10}
\emline{82.33}{75.67}{11}{107.67}{75.67}{12}
\emline{82.00}{75.33}{13}{107.67}{65.00}{14}
\put(21.00,86.01){\makebox(0,0)[cc]{$|\psi_0\rangle$}}
%\put(21.00,88.01){\makebox(0,0)[cc]{}}
\put(69.00,100.00){\makebox(0,0)[cc]{$|b\rangle$}}
\put(69.00,75.33){\makebox(0,0)[cc]{$|b'\rangle$}}
\put(75.33,100.33){\makebox(0,0)[cc]{$\bigl( {1\over 3}\bigr)$}}
\put(75.00,75.34){\makebox(0,0)[cc]{$\bigl( {2\over 3}\bigr)$}}
\put(111.33,111.00){\makebox(0,0)[cc]{$|\psi_f\rangle$}}
\put(111.33,99.67){\makebox(0,0)[cc]{$|\psi_{f}'\rangle$}}
\put(111.33,89.67){\makebox(0,0)[cc]{$|\psi_{f}''\rangle$}}
\put(112.33,75.33){\makebox(0,0)[cc]{$|\psi_{f}'\rangle$}}
\put(111.67,64.00){\makebox(0,0)[cc]{$|\psi_{f}''\rangle$}}
\put(117.67,111.00){\makebox(0,0)[cc]{$\bigl( {1\over 9}\bigr)$}}
%\put(113.00,-10.00){\makebox(0,0)[cc]{$|\psi_{f}''\rangle$}}
\put(2.00,86.33){\makebox(0,0)[cc]{(b)}}
\emline{82.00}{75.33}{15}{108.33}{84.67}{16}
\put(117.67,84.33){\makebox(0,0)[cc]{$(0)$}}
\put(111.33,84.33){\makebox(0,0)[cc]{$|\psi_f\rangle$}}
\put(70,31.00){\makebox(0,0)[cc]
{\normalsize (a): A measurement of observable $A$ is 
performed at time $t_1$. The numbers in parentheses}}
\put(70,26.00){\makebox(0,0)[cc] {\normalsize indicate
the fraction of particles selected in the given
state at each measurement.}}
\put(48,21.67){\makebox(0,0)[cc]
{\normalsize (b): A measurement of observable $B$ is 
performed at time $t_1$.}}
%\vector(27.33,50.00)(112.00,50.00)
\put(112.00,50.00){\vector(1,0){0.2}}
\emline{27.33}{50.00}{17}{112.00}{50.00}{18}
%\end
\put(73.33,44.33){\makebox(0,0)[cc]{$t_1$}}
\put(28.33,45.00){\makebox(0,0)[cc]{$t_0$}}
\put(109.67,45.00){\makebox(0,0)[cc]{$t_f$}}
\large
\end{picture}
 
\newpage
 Of those systems, roughly $1/3$ will be found in box A
 and $2/3$ will be found to be not
in box A. Subsequently, when we perform the 
post-selection measurement of $Q_f$, the distribution will be as follows: 
of those systems found at time $t_1$ in box A, $1/3$ will be 
post-selected in the state $\psif$.  However, {\it none}
of the particles that 
were found to be {\it not} in A can be post-selected in 
state $\psif$, since the state $|a'\rangle$ corresponding to
``not in box A'' is orthogonal to $\psif$. 
Thus the actual process occurring in this experiment is 
one in which roughly $1/9$ of the pre-selected particles will be 
post-selected; and all the particles that are post-selected will be 
{\it guaranteed} to be ones that were found in box A at time
 $t_1$. 
  
If we consider an experiment
 in which the observable $B$ is measured at 
time $t_1$, we observe exactly the same statistics but with the roles of 
observables $A$ and $B$ interchanged (Figure 2b). In each case,
 any particles which end up post-selected are 
ones which {\it could not have been in 
any box except the one which was opened} (be it A or B).
Thus, we see that a necessary (but not sufficient) condition for post-selection of a
particular particle X via a measurement of (for 
instance) $A$ is that 
particle X was found in box A at time $t_1$. 
Since the same particle X's being found in box A 
and being found in
box B at the same instant of time are mutually exclusive
states of affairs, for purely physical reasons it is
clearly incorrect to say of any such particle X
that it had a probability 1 of being 
found in box B at time $t_1$.

To make a claim about the elements of reality of
an individual system, we have to consider
the physical situation involved in an individual
run of the experiment.
But in each run, we have to make a choice 
as to measure A or
B. In the cases that we choose to measure A, all successfully
post-selected particles had to be found in box A at $t_1$,
and {\it mutatis mutandis} for a measurement of B.
 This means that it 
is not valid to say of any {\it individual particle}, ``If in the intermediate time
 it was searched for in box A, it has to be found there with probability 
one, and if, instead, it was searched for in box B, it has to be found 
there too with probability one...''  (Vaidman, 1996).
	
	The same argument applies to the generalized example of a 
particle in $N+1$ boxes as discussed by Aharonov and Vaidman (1991). 
Aharonov and Vaidman say of this example, ``in spite of the fact that we have only one 
particle in the above situation, we find this particle with probability 
one in any one of the first $N$ boxes'' (1991, p. 2318).  However, as shown 
above, this statement is inaccurate in the sense that the property of 
being with certainty in any one of $N$ boxes (depending on which 
one is opened) {\it cannot apply to the same individual particle} 
in any given run of the experiment.  Thus these `elements of reality,' 
as defined by Vaidman, are not really properties of individual 
systems but apply only to ensembles.

	The fallacy of attributing these ``peculiar'' properties to a 
single particle can be seen as arising directly from the counterfactual reading of 
the ABL rule. As noted above, in any given run of the experiment
in a which a given particle X is post-selected, we can measure  
only one of the two observables $A$ and $B$. Once we have chosen one of 
these, say $A$, it 
is erroneous to apply a counterfactual reading 
of the ABL rule to particle X
with respect to a measurement of observable $B$ which 
has not actually occurred in the process of post-selecting
that particle. (As noted in section 1, this point
is further elaborated upon in the Appendix, which
discusses conditions for validity of the counterfactual
ABL usage in terms of Griffiths' consistent histories
formalism.)\footnote{\normalsize cf. Griffiths (1984), (1996), 
(1998).}
	
{\bf 4. Critique of Vaidman's counterarguments}

In his most recent paper, Vaidman (1998) presents counterarguments
to some recent refutations
of the counterfactual usage of the ABL rule. 
Vaidman claims that the refutations are based on confusion
about the evaluation of counterfactual statements in quantum
theory, especially with respect to its indeterminism; and about
the role of time symmetry in counterfactual statements. However, none
of these counterarguments succeed in identifying any flaw in
the refutations, as will be shown below.

 Vaidman starts by noting that ``there is a general
philosophical trend to consider counterfactuals to be asymmetric
in time'' (1998, p.2). He then quotes 
Lewis: ``I believe that indeterminism is neither
necessary nor sufficient for the asummetries I am discussing.
Therefore I shall ignore the possibility of indeterminism in
the rest of this paper, and see how the asymmetries might
arise even even under strict
determinism.'' (Lewis 1986, p.37) 
Vaidman states that, ``in contrast to this opinion'', he believes that 
``indeterminism is crucial for allowing non-trivial time-symmetric
counterfactuals, and that Lewis's and other general philosophical
analyses are irrelevant for the issue of counterfactuals
in quantum theory.'' 

This opening statement is a very
curious one for several reasons. It starts by making an observation
that time asymmetry is usually assumed in theories of counterfactuals,
while making no attempt to show that such theories {\it depend} in
any way on time asymmetry--the latter being the crucial
consideration.  The mere fact that time asymmetry may be
a metaphysical presupposition or prejudice falls far short of
showing that classical theories of counterfactuals are inimical
or irrelevant to time-symmetric counterfactuals. 
So the observation about time-asymmetry preferences
in no way casts doubt on the soundness or appropriateness
of theories such as Lewis'  or Stalnaker's (1968) for
analyzing time-symmetric counterfactuals.
 
Now to the question of indeterminism. The sketch of an argument
by Vaidman,
given above, is incomplete at best. 
It appears to take Lewis' above-quoted statement in favor of 
the sufficiency of determinism for time asymmetric
counterfactuals
as being contrary to Vaidman's view in favor of the necessity of 
indeterminism for time {\it symmetric} counterfactuals, 
which it is not. Vaidman then summarily takes
the same argument by Lewis as an indication {\it against the applicability of Lewis'
theory to indeterministic or time-symmetric situations}.

However, none of these negative assertions about Lewis' theory
has been supported by sound, or even anything 
approaching complete, arguments.
In particular, it has not been shown that Lewis' theory
is intrinsically inapplicable either to 
 indeterminism
or to time symmetry.\footnote {\normalsize A good indication that 
Lewis in no way presupposes determinism in his theory
is the following statement, (Lewis 1973, p. 75):
``Suppose that the laws prevailing at a world {\it i}
are deterministic as we used to think the laws
of our own world were.''} Thus, it is specious
to suggest that any theory of counterfactuals
that considers deterministic and/or time asymmetric
situations is automatically disqualified for use
with indeterministic and/or time-symmetric situations.
 Vaidman's conclusion that
``Lewis's and other general philosophical
analyses are irrelevant for the issue of counterfactuals
in quantum theory'' would appear to be a completely groundless dismissal
of a perfectly general and sound counterfactual theory.

Furthermore, Vaidman then goes on to claim that ``the key
questions in [Lewis' and similar theories]
are related to [the antecedent] $\cal A$....Do we
need a `miracle' (i.e., breaking the laws of physics)
for $\cal A$?...'', and asserts that ``miracles''
are ``the main topic of discussion on counterfactuals
in general philosophy.'' (1998, p.2) Again, this is simply
incorrect. In his seminal book {\it Counterfactuals}, Lewis
spends three pages, out of a total of 142, 
discussing miracles (1973, pp. 75-77); this is in connection
with a discussion of the importance of laws of nature in determining
the closeness of worlds under the relevant
similarity relation. Thus, despite Vaidman's
suggestions to the contrary, Lewis' theory is perfectly applicable
in an indeterministic universe where miracles are irrelevant.

 Vaidman goes on to propose a definition for time-symmetric 
counterfactuals in quantum theory. His
definition is based on a proposed similarity relation
which ``fixes'' the pre- and post-selection outcomes. A detailed
critique of this
definition has already been given in Kastner (1998);
these arguments will not be repeated here except to
note that the ``fixing'' requirement is highly
problematic and amounts to proposing an unphysical
similarity relation. 

In this paper, we discuss an additional problem
with Vaidman's proposal for a time-symmetrized counterfactual:
 a problem with the syntax of the definition which reflects
a confusion between the non-counterfactual and
counterfactual usage of the ABL rule. 

Vaidman's proposed definition
is: \newline \indent
``If it were that a measurement of an observable $A$
 has been performed at time $t_1$, \hbox{$t_0 < t_1 < t_f$}, then
the probability for $A = a_i$ would be equal to $p_i$,
provided that the results of measurements performed
on the system at times $t_0$ and $t_f$ are
 fixed.''\footnote{I have slightly changed the notation for
consistency with that used in this discussion
and in the Appendix. In the original quote,
Vaidman uses $t_1$ instead of $t_0$,
$t$ instead of $t_1$, and $t_2$ instead of
$t_f$.} (1998, p.5)\qquad (*)

The above definition incorporates 
a strange and awkward mixing
of tenses: ``If it {\it were} that...''(the subjunctive tense)
 juxtaposed with
``a measurement of an observable $A$ {\it has been} performed...'' (the past
perfect tense). This muddling
of tenses suggests that the definition attempts to
strike a `middle ground' between two distinct usages:
(a) ``If observable $A$ {\it was} measured, then the probability
for $A = a_i$ (at time $t_1$) {\it was} equal to $p_i$.'' 
This is the non-counterfactual 
(i.e., material conditional) usage of the
ABL rule, and it is correct, as discussed in Kastner (1998).
(b) ``If observable $A$ {\it had been} measured 
(instead of some other observable which was actually 
measured), then the
probability  for $A = a_i$ (at time $t_1$) {\it would have been}
$p_i$.'' This is the {\it bona fide}
subjunctive conditional, or counterfactual,
usage, and it is generally incorrect, as discussed in Kastner (1998).

Consider the two usages (a) and (b) in the context
of an actual experimental situation. If observable
$A$ was {\it actually} measured at time $t_1$, then usage (a) applies nontrivially
and usage (b) reduces to the case of a counterfactual
with true antecedent, or a material conditional
 (see Lewis 1973, p. 26); in other words it becomes logically
equivalent to (a). If, on the other hand, observable $A$
was {\it not} actually measured, but instead some other (noncommuting)
observable $B$, then usage (a) is still correct but
now applies only vacuously (i.e., proposition (a) is vacuously true).
Meanwhile, usage (b) becomes a {\it bona fide} counterfactual
with false antecedent; this usage is now incorrect, and proposition (b)
is false.

Once it is admitted that some definite observable (perhaps
the trivial observable $I$ corresponding to no measurement)
was {\it actually} measured in the selection of any given
system--regardless of whether or not we are privy
to that information--we are forced to choose between the two 
situations described above.
 If observable $A$ was actually measured at time $t_1$,
then (a) is equivalent to (b) and they both apply; if observable $A$
was {\it not} actually measured at time $t_1$, then 
(a) is still (vacuously) correct but (b) is not
(except in certain special cases, as discussed in Kastner (1998),
Cohen (1995)).
Thus Definition (*), as it stands, is grammatically incorrect in
a way that reflects its lack of clarity and rigor
with respect to the physically crucial point concerning
which measurement has {\it actually} taken place.

{\bf 5. Conclusion}

It has been argued that an apparent paradox proposed
by Aharonov and Vaidman (1991), and further
amplified by Vaidman (1996), to illustrate peculiarities of time-symmetric
quantum systems is not a true paradox, but arises
from an invalid counterfactual usage of the ABL rule.
The paradox consists in the apparent assignment of
mutually exclusive properties to a system; however
it is resolved by noting that these properties can
never be simultaneously attributed to the same individual system.
A connection is made between this problem and the consistent histories
approach of Griffiths.
In addition, some counterarguments by Vaidman against
refutations of the counterfactual usage of the ABL rule
are analyzed and shown to be ineffective. 

{\bf 6. Appendix}

The consistent histories (henceforth `CH') approach pioneered by Griffiths
(1984, 1996, 1998) has been widely discussed in connection
with the problem of assigning properties to quantum
systems independent of measurement. In particular, Cohen (1995, pp. 4376-7)
gives a concise summary of the basic features of the
formulation in that context. In this Appendix we will very briefly review
the fundamental features of CH in order to relate it to the
ABL rule. Readers desiring a more complete exposition of
CH may wish to refer to the Griffiths and/or Cohen
references noted above.

A ``history'' as defined by Griffiths is a series
of events 

$$D \to E_1 \to E_2 \to \dots \to F, \eqno (5)$$ 
occurring at times $t_0, t_1, \dots t_f$, with the
subscripts on the events denoting the time of their
occurrence; events D and F represent the initial and
final events occurring at $t_0$ and $t_f$, respectively.

A particular history is considered as a member of 
a family of histories associated with the ``events sets''
$[E^{\alpha}_i]$, where $[E^{\alpha}_i]$ is a set of 
orthogonal projections, i.e.,

$$E^{\alpha}_k E^{\beta}_k = \delta_{\alpha\beta} E^{\alpha}_k\eqno(6)$$

comprising a decomposition of the identity:

$$1 = \sum_{\alpha} E^{\alpha}_k.\eqno(7)$$

For the special case in which we are interested, that is,
three time indices ($t_0 < t_1 < t_f$) and zero Hamiltonian,
a given history is considered to be a {\it consistent history}
if and only if, for all $ \alpha \ne \beta $ the histories 
in its associated event sets satisfy the condition:

$$ Re Tr ((DE^{\alpha}F)^{\dag} \ DE^{\beta}F) = 0.\eqno (8)$$

This condition ensures that the probabilities of disjoint individual
histories comprising the family are additive, thus 
disallowing quantum mechanical `interference'
between mutually exclusive histories. Families of
histories satisfying condition (8) are called
``consistent families'', or ``frameworks''.

The probability of a given consistent history $Y$, 
$$Y= D \wedge E^{\alpha} \wedge F\eqno (9)$$
is then given by 
$$P(Y) = Tr((DE^{\alpha}F)^{\dag} DE^{\alpha}F) = 
Tr (E^{\alpha}DE^{\alpha}F),\eqno(10)$$
using the fact that projection operators are idempotent
and self-adjoint, and that the trace is invariant
under cyclic permutations.
Since the above now behaves exactly as an ordinary classical
probability, this result can be extended in the usual way
to conditional probabilities via Bayes' rule, i.e.:

$$P(E^{\alpha}\vert D\wedge F) = {P(D\wedge E^{\alpha} \wedge F) 
\over P(D \wedge F)} \eqno(11)$$

$$= Tr(E^{\alpha}DE^{\alpha}F) /Tr(DF).\eqno(12)$$

The consistent histories formulation applies to a closed
system usually taken to be a composite of quantum system S
and measuring apparatus M, with associated Hilbert
space $\cal H = \cal S \otimes \cal M$.
\def\QD{\vert D\rangle}
\def\QF{\vert F\rangle}
In order to relate the conditional probability in (11)
to the ABL rule, consider a typical experiment in which
system S is preselected in state $\QD$ and post-selected
in state $\QF$. In accordance with the notation of
Griffiths, we denote the projection operators associated
with quantum states simply by the letters labeling the state.
Apparatus M, which measures the post-selection observable, 
starts out at time
$t_0$ in a ready state $M_0$. At time $t_1$, the 
apparatus remains untriggered but we consider a framework
in which S has some value $C_k$ associated with an arbitrary
observable $C$ defined over the system Hilbert space $\cal S$.
At time $t_f$, apparatus M has been triggered in state
$M_F$ corresponding to finding the system in state $F$.

Thus the history analogous to (5) is in this case:
$$D \otimes M_0 \to C_k \otimes M_0 \to 
F \otimes M_F,\eqno(13)$$

We can now make the connection with the ABL rule,
which (in its counterfactual form) essentially asks: 
What is the probability that the system
is in state $C_k$ at time $t_1$, given that
it was preselected in state $D$ and post-selected in state $F$?
\footnote{\normalsize The non-counterfactual usage corresponds to
the question: Given that a measurement of observable $C$ is
performed at $t_1$ on the pre- and post-selected system, what is
the probability that the measurement outcome is $C_k$?}

In the Griffiths formalism, this probability is given by:

$$  P(C_k\vert (D\otimes M_0) \wedge (F \otimes M_F))=
{Tr[(D\otimes M_0)(C_k\otimes I)(F \otimes M_F)(C_k\otimes I)]
\over Tr[(D\otimes M_0)(F \otimes M_F)]}\eqno(14)$$

$$={Tr(D C_k F C_k) Tr(M_0 M_F)\over Tr(DF) Tr(M_0 M_F)}= 
{\sum_i \langle C_i \vert D C_k F C_k \vert C_i\rangle
\over \sum_i \langle C_i \vert D F \vert C_i\rangle}
= {\langle C_k \vert D C_k F \vert C_k\rangle\over
\sum_{i,j} \langle C_i \vert D \vert C_j\rangle
\langle C_j \vert F \vert C_i \rangle}\eqno(15)$$.

Note that (15) is only equivalent to the ABL rule
if we further assume that the history Y is consistent,
\footnote{\normalsize This discussion assumes we
are considering a counterfactual measurement, i.e.,
probabilities associated with outcomes of an observable $C$
that has not actually been measured. If $C$ is actually
measured at time $t_1$, then (according to the
orthodox interpretation) the interference terms 
corresponding to $i \ne j$ vanish upon measurement.}
i.e., that for $i \ne j$,
$$ Re (\langle C_i \vert D \vert C_j\rangle
\langle C_j\vert F \vert C_i\rangle) = 0.\eqno(16)$$

Applying that condition, we then can say:
$$  P_{CH}(C_k\vert D\otimes M_0 \wedge F \otimes M_F)=
{\langle C_k \vert D C_k F \vert C_k\rangle\over
\sum_{i} \langle C_i \vert D \vert C_i\rangle
\langle C_i\vert F \vert C_j\rangle} 
= {\vert \langle D \vert C_k\vert F \rangle \vert^2
\over \sum_i {\vert \langle D \vert C_i\vert F \rangle \vert^2}},
\eqno(17)$$
which is the ABL rule.
Thus, the ABL rule can be obtained as a special case of
the consistent histories approach. 

The counterfactual usage
of the ABL rule is equivalent to asserting that histories
associated with an observable that was not
actually measured at time $t_1$ may be meaningfully
added to the event set $\{D,[E^{\alpha}],F\}$. In the three-box example, 
the two observables that could be measured at $t_1$ are $A$
and $B$. The counterfactual usage thus corresponds to the
set of histories
$$ \psi_0 \to [A^{\alpha}, B^{\alpha}] \to \psi_f \eqno(18)$$

\noindent where\vskip 2pc

\noindent $A^1 = a$,\newline
 $A^2 = b + c$,\newline
 $B^1 = b$,\newline
and $B^2 = a + c.$\vskip 2pc

However, (18) is not a consistent family, as can be
seen by applying the consistency condition (8) or (16). For
$E^{\alpha} = a$ and $E^{\beta}= b$ we find the nonvanishing result:

$$ Re Tr \Big[ (\psi_0 \ a \ \psi_f)^{\dag}\ 
 \psi_0\  b \ \psi_f\Big] =
Re \Big[ \langle a \vert \psi_0 \vert b\rangle
\langle b\vert \psi_f \vert a\rangle) \Big] = 
{1\over 9} \ne 0. \eqno(19)$$ 

Since condition (16) fails, we cannot
use the ABL rule to calculate the probability of any
particular value of either $A$ or $B$ at time $t_1$,
but instead must use (15). As Griffiths (1984, 1996, 1998) has shown,
only if the family of histories is consistent can we
combine probabilities in the classical way so as to make 
inferences such as ``If I had measured $B$ instead of $A$ at $t_1$,
particle X would have been in box B''. 

According to Griffiths, it is ``not meaningful''
to consider together the probabilities of histories that
are not consistent, i.e., which belong to different frameworks.\nolinebreak
\footnote{\normalsize Cf. Griffiths (1998, p. 1607).}
While terms such as ``not meaningful'' might be criticized on
the same basis as has been classical positivism (i.e., it has long
been recognized that dismissals by positivists of certain
statements as ``meaningless'' have been untenable 
\footnote{\normalsize Cf. Popper (1959, pp. 35-6).}),
Griffiths' admonition not to combine probabilities of
histories belonging to inconsistent families has important physical
content, as can be seen in reference to the three-box example.
Specifically, in this example,  
the proscription against combining those probabilities obtained
by applying the ABL rule first to a measurement of observable $A$
and then again to a measurement of observable $B$ corresponds 
physically to
the fact that the associated properties (being in box A or
being in box B) cannot be possessed by the same individual
particle.

{\bf Acknowledgements}

The author gratefully acknowledges valuable correspondence
and/or discussions with J. Bub,
R. B. Griffiths, and J. Malley.

{\bf References}

\noindent Aharonov, Y, Bergmann, P.G. , and Lebowitz, J.L. (1964), 
`Time Symmetry in the Quantum Process of Measurement,' 
{\it Physical Review B 134}, 1410-16.\newline
Aharonov, Y. and Vaidman, L. (1991), `Complete Description
of a Quantum System at a Given Time,'
{\it Journal of Physics A 24}, 2315-28.\newline
Albert, D.Z., Aharonov, Y., and D'Amato, S. (1986),
``Comment on `Curious Properties of Quantum Systems
Which Have Been Both Preselected and Post-Selected,''
{\it Physical Review Letters 56}, 2427.\newline
Cohen, O. (1995), `Pre- and postselected quantum systems, 
counterfactual 
measurements, and consistent histories, '
{\it Physical Review A 51}, 4373-4380.\newline
Griffiths, R.B. (1984), `Consistent Histories and the
Interpretaton of Quantum Mechanics,' {\it Journal of Statistical 
Physics 36}, 4373.\newline
Griffiths, R.B. (1996), `Consistent Histories and Quantum
Reasoning,' {\it Physical Review A 54}, 2759-2774.\newline
Griffiths, R.B. (1998), `Choice of Consistent
Family, and Quantum Incompatibility,' {\it Physical Review 
A 57}, 1604-1618.\newline
Kastner, R.E. (1998), `Time-Symmetrized Quantum Theory,
Counterfactuals, and ``Advanced Action,''' to be published,
{\it Studies in History and Philosophy of Modern Physics};
quant-ph/9806002.\newline
Miller, D.J. (1996), `Realism and Time Symmetry in 
Quantum Mechanics,' {\it Physics Letters A 222}, 31.\newline
Popper, Karl R. (1959), {\it The Logic of Scientific Discovery},
London: Routledge.\newline
Sharp, W. D. and Shanks, N. (1993), `The Rise and Fall of 
Time-Symmetrized 
Quantum Mechanics,' {\it Philosophy of Science 60}, 
488-499.\newline
Stalnaker, R. (1968), `A Theory of Conditionals,' 
in {\it Studies in Logical Theory}, edited by N. Rescher. 
Oxford: Blackwell.\newline
Vaidman, L. (1996), `Weak-Measurement Elements of Reality,' 
{\it Foundations of Physics 26}, 895-906.\newline
Vaidman, L.(1998), `Time-Symmetrized Counterfactuals
in Quantum Theory,' preprint, quant-ph/9802042.
\end{document}